# SAMPLE NO IEE 15070003 DESCRIPTION: AMMOCOETES OF *LETHENTERON CAMTSCHATICUM* (TILESIUS, 1811) (PETROMYZONTIDAE, PETROMYZONTIFORMES) FROM MALAYA KHUZI RIVER (SAKHALIN)

A.V. KUCHERYAVYY, E.A. KIRILLOVA

*Laboratory of the Lower Vertebrate Behavior, Institute of Ecology and Evolution, Russian Academy of Sciences, Leninskii prospect, 33, Moscow, Russia*

*Contact e-mails: scolopendra@bk.ru; lisaveta_m@mail.ru*

Key words: Arctic lamprey; larvae; downstream migrants; measurements; counts; age; gonad development

**Abstract –** Paper provides biological description of both males and females of Arctic lamprey, *Lethenteron camtschaticum* ammocoetes from one of Sakhalin rivers, Malaya Khuzi. It briefly describes stages of development of the larvae gonads and cases of infantilism and transverse.

## Introduction

Expanding Collection of lampreys at the Laboratory of the lower vertebrate behavior, Institute of Ecology and Evolution, Russian Academy of Sciences has led to the idea of generally available Database creation. By the present time this base is being developed, its goal is to help researchers to receive reliable information on lampreys, various stages of their ontogenesis, distribution, classification and features etc. The database will be connected to several Internet resources, like Freshwater Ecoregions of the World (feow.org) and other. The samples descriptions, i.e. this one, are to be published on the open-access sources, to get referenced and cited at the sample brief description on the Database.

## Material and Methods

Specimen body proportions were measured as described in [1; 2]. Stage of gonad development estimated acc. to [3]. Measurements were taken on specimens preserved in 4% solution of formaldehyde. Tissues were taken from the middle of the ammocoete body and studied acc. to the standard protocol for gonads preparation in Ehrlich's hematoxylin-eosin. Sections thickness was 4-5 µm. Measurements and counts abbreviations: FES, freshwater ecosystem [4]; first stage of ovary development, FSO; second stage of ovary development, SSO; transverse gonad, TNS; infernal gonad, INF; total body length, ab; body wet weight, m; number of trunk myomers, n.mio; maximal body depth, dc; preocular length, ag; prebranchial length, ae; branchial length, ej; head length aj; trunk length jl; tail length lb; distance between $1^{st}$ and $2^{nd}$ branchial openings, $e_1e_2$; length of oocyte, Ø; condition factor, as $\log_{10}(ab/m)$, CF. Specimens are combined into a group in case if at least 6 representatives have the same signs of aggregation, otherwise description is given for each individual.

## Label information

| | | | |
|---|---|---|---|
| **Sample ID** | | 15070003 | |
| **Det. as** | | *Lethenteron camtschaticum* (Tilesius, 1811) | |
| **Stage of development** | | Ammocoetes | |
| **Sample size** | | *n*=66 | |
| **Coll. by:** | Kirillova; Kirillov | **Det. by:** | Kucheryavyy |
| **Date of coll.** | Jun, 3 – Jul, 16 2015 | **Place** | Malaya Khuzi |
| **N.B.!** | Specimens were caught during downstream migration survey at night time. Survey site was at the right side of the river; depth, 0.8–0.5 m; water temperature, 2.1–15.6 °C; velocity, 3-0.9 m/s. Distance from the sea 1.1 km. Downstream migrants were caught by a cone trap made of bolting silk no. 10. The trap with square inlet with a 0.5 m side was set on the rope, attached to a line stretched across the river channel. Length of the trap, 2.5 m. A plastic bucket with a removable lid was attached to the top of the cone trap. Samples were taken every 30 minutes, exposition of the trap, 0.5–5 min. | | |

## Geographic description

| | | | |
|---|---|---|---|
| **Sublocality** | Lower course of mountain-type river | **Locality** | Fig. 1 |
| **River** | Malaya Khuzi | **Topography** | click here |
| **Cluster** | Sakhalin | **FES** | 641, click here |
| **Subgroup** | Sakhalin – Eastern Sikhote Alin | **Group** | Okhotsk and Primorye |
| **Marine Basine** | Sea of Okhotsk | **Ocean** | Pacific |
| **Subprovince** | Eastokhotian | **Province** | Okhotian mountain |
| **Subregion** | Easteurosiberian | **Region** | Eurosiberian taiga (boreal) |

## Grouping criteria

| Sign of aggregation | Applicability | Comments |
|---|---|---|
| **Sex** | YES | 3 groups (♀, ♂, unspecified) |
| **Feature** | Gonad development | 5 groups (Table 1) |

## Morphology

*Sample description*. Young postmetamorphic specimens, abs. Adults, abs. Sample consists of 66 ammocoetes with total length of 31-150 mm. For morphological analysis and gonads histology 26 specimens are taken. Gonads analysis discovered three sexual groups, i.e. males (group 1), females (groups 2-3) and unspecified (groups 4-5). In females there were specimens with ovaries of the first (group 2) and second (group 3) stages of development. In the sexually unspecified group there were three specimens of infernal stage (group 5) and one (group 4) transverse (Table 1).

*Sample analysis and conclusion.* Ammocoetes have a typical eel-like shape with 2 dorsal fins. Paired fins lack. Body wet weight in specimens 50-150 mm, 0.2-5.0 g. Body proportions, as percentage of total length: prebranchial length, 6.5-11.0; branchial length, 18.1-26.3; trunk length, 45.5-62.6; tail length, 15.8-36.3; eye length, unrec.; disc length, abs. Urogenital papilla, abs. Trunk myomeres, 63-76. Dentition, abs. Velar tentacles, abs. Caudal fin shape, spade-like. Oral fimbriae, abs. Oral papillae, abs. Detailed information, Table 1.

Maximal body length, 150 mm; maximal body weight, 5.0 g; maximal age estimated as four years (Fig 2, 3). 4 age classes (I-IV). General sample length-weight parameters described as the equation, $m=0.049e^{0.0313ab}$; $R^2=0.9752$ (Fig.4).

Field works and sample study are supported by the RNF 14-14-01171

**Table 1.** Detailed description of *Lethenteron camtschaticum* ammocoetes from Malaya Khuzi, collected on June – July, 2015

| Feature | Males (n=8) | Females | | | | Unspecified | | | |
|---|---|---|---|---|---|---|---|---|---|
| | | FSO, 1 | FSO, 2 | FSO, 3 | SSO (n=11) | TNS, 1 | INF, 1 | INF, 2 | INF, 3 |
| Group | 1 | 2 | | | 3 | 4 | 5 | | |
| ***ab*, mm** | 98-141 / 128.5±15.56 | 105 | 105 | 94 | 87-143 / 111.9±17.69 | 134 | 127 | 147 | 98 |
| ***m*, g** | 1.0-3.8 / 2.74±1.02 | 1.3 | 1.2 | 0.9 | 4.8-5.8 / 5.51±0.31 | 3.1 | 3.0 | 3.8 | 1.2 |
| ***n.mio*** | 67-76 / 70.5±2.93 | 67 | 63 | 66 | 63-73 / 68.50±3.38 | 67 | 66 | 75 | 66 |
| In %, *ab* | | | | | | | | | |
| ***dc*** | 4.8-5.8 / 5.41±0.36 | 5.3 | 5.3 | 4.7 | 4.8-5.8 / 5.51±0.31 | 5.6 | 5.7 | 5.1 | 5.2 |
| ***ag*** | 2.3-3.0 / 2.69±0.19 | 2.7 | 2.7 | 2.8 | 2.2-3.0 / 2.77±0.25 | 2.8 | 2.5 | 2.4 | 3.4 |
| ***ae*** | 7.0-11.0 / 8.10±1.24 | 7.7 | 8.5 | 8.7 | 6.6-9.9 / 8.37±1.24 | 7.2 | 7.9 | 6.7 | 8.2 |
| ***ej*** | 12.6-15.2 / 13.79±0.82 | 13.8 | 13.8 | 13.7 | 11.8-16.4 / 14.03±1.35 | 13.5 | 14.6 | 13.1 | 13.8 |
| ***aj*** | 20.4-26.2 / 21.88±1.86 | 21.4 | 22.3 | 22.4 | 18.3-26.3 / 22.40±2.43 | 20.7 | 22.6 | 19.8 | 22.0 |
| ***jl*** | 50.0-54.6 / 51.62±1.69 | 49.5 | 48.6 | 47.3 | 45.5-62.6 / 52.09±5.09 | 49.3 | 55.9 | 50.3 | 50.0 |
| ***lb*** | 23.8-28.3 / 23.48±1.57 | 29.0 | 29.2 | 30.2 | 15.8-36.3 / 25.52±6.44 | 30.1 | 21.5 | 29.9 | 28.0 |
| ***$e_1e_2$*** | 2.0-2.9 / 2.27±0.36 | 1.9 | 1.9 | 2.0 | 1.8-2.6 / 2.19±0.30 | 1.9 | 3.1 | 2.0 | 1.9 |
| ***Ø*** | - | 20.3 | 16.0 | 29.0 | 50.2-88.5 / 70.64±12.20 | - | - | - | - |
| ***CF*** | 1.6-2.0 / 1.70±0.15 | 1.9 | 1.9 | 2.0 | 1.5-2.0 / 1.81±0.14 | 1.6 | 1.6 | 1.6 | 1.9 |

Note: in groups 1 and 3 above the line, limits; under the line M±SD

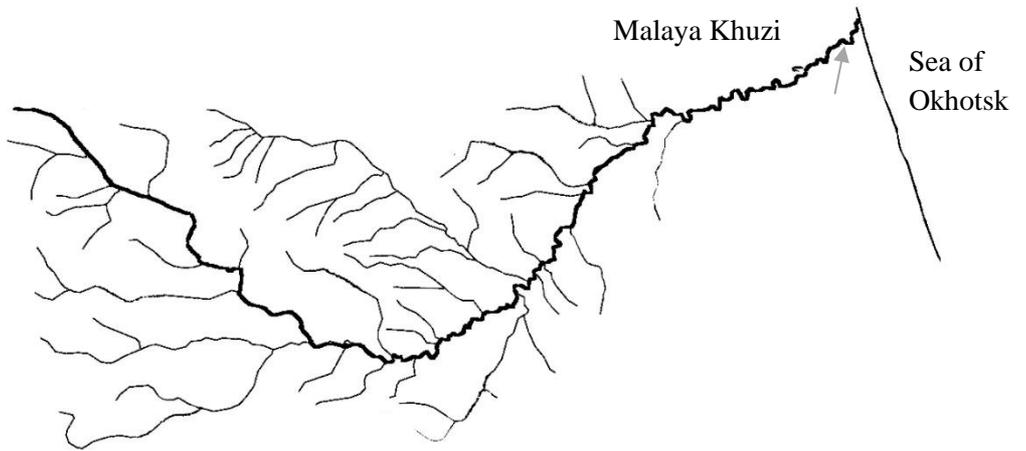

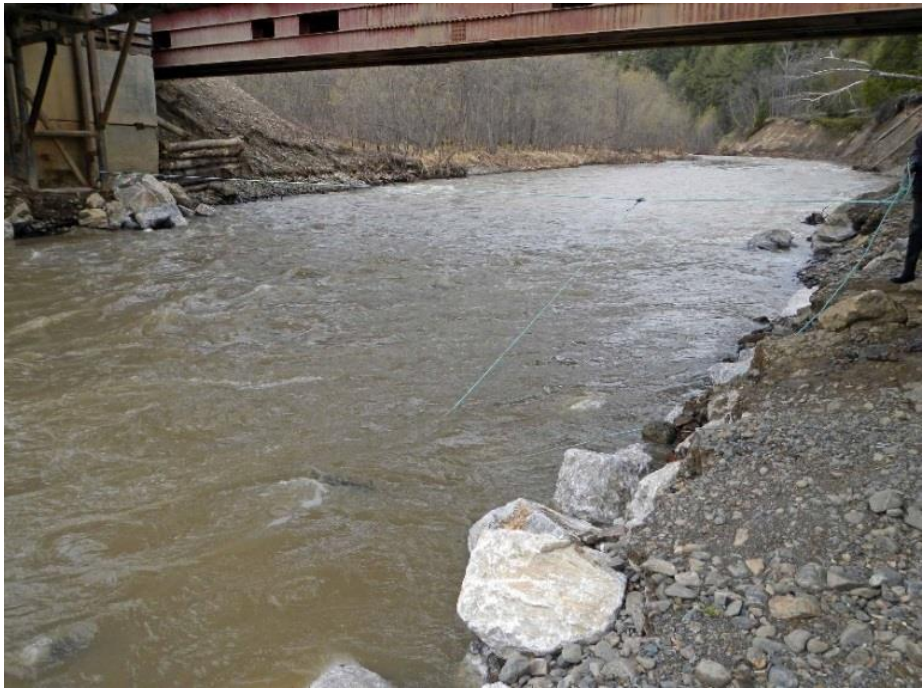

**Figure 1.** Site of the sample collection of *Lethenteron camtschaticum* ammocoetes in June – July, 2015

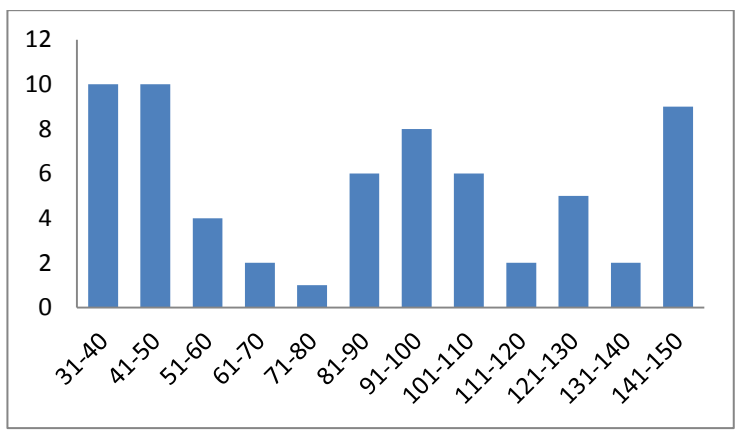

**Figure 2.** Size composition of *Lethenteron camtschaticum* ammocoetes sample from Malaya Khuzi, collected on June – July, 2015. Abscises, total body length, mm; ordinates, number of specimens.

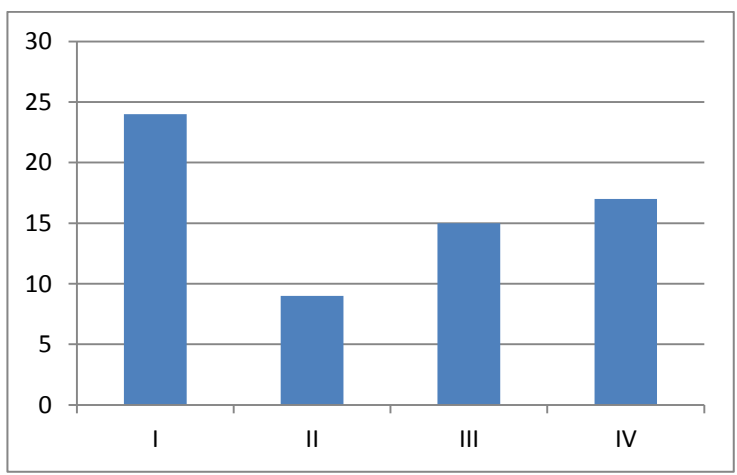

**Figure 3.** Age composition of *Lethenteron camtschaticum* ammocoetes sample from Malaya Khuzi, collected on June – July, 2015. Abscises, estimated age, y; ordinates, number of specimens.

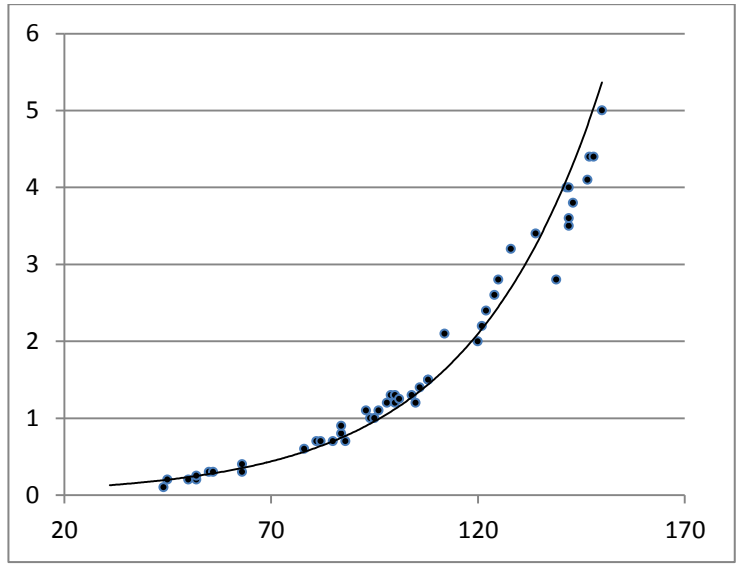

**Figure 4.** Body length – body weight correlation for *Lethenteron camtschaticum* ammocoetes sample from Malaya Khuzi, collected on June – July, 2015. Abscises, total body length, mm; ordinates, body wet weight, g.